# Longitudinal jitter analysis of linear accelerator electron gun

L. Mingshan,[1] P. Shilun,[1] and M. Iqbal[1,2a)]

[1]Institute of High energy Physics, Chinese Academy of Sciences, Beijing 100049, China
[2]Centre for High Energy Physics, University of the Punjab, Lahore 45590, Pakistan

We present measurement and analyses of longitudinal timing jitter of Beijing Electron Positron Collider (BEPCII) linear accelerator electron gun. We simulated longitudinal jitter effect of the gun using PARMELA about beam performance including beam profile, average energy, energy spread, longitudinal phase of reference particle and XY emittance. The maximum percentage difference of the beam parameters are calculated to be; 100%, 13.27%, 42.24%, 7.79% and 65.01%, 86.81%, respectively due to which the bunching efficiency is reduced to 54%. The simulation results are in agreement with test and are helpful to optimize the beam parameters by tuning the trigger timing of the gun during the bunching process.

Electron gun timing longitudinal jitter is fatal not only for electron beam performance but also for positron yield in routine operation of Beijing Electron Positron Collider (BEPCII) Linac, which has been observed many times practically. Jitter of the gun in the transverse direction caused by gun high voltage has already been reported elsewhere[1]. The present study is about beam bunching performance caused by jitter of gun timing in the longitudinal direction. This trigger is caused due to many factors including its electronic components, temperature, etc. The threshold value for the longitudinal jitter is less than 50ps to ensure expected bunching results, in the pre-injector[2].

The sub-harmonic bunching (SHB) system of BEPCII Linac pre-injector is composed of electron gun, two sub-harmonic bunchers (SHB1 & SHB2), one 4-cell travelling wave buncher and a standard 3-m long accelerating structure [3-4] as shown in Fig. 1.

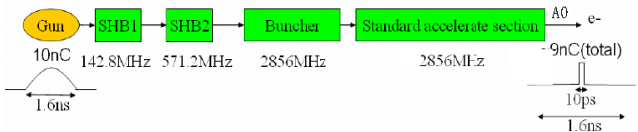

FIG.1. (Color online). Schematic of the pre-injector

The electron gun[5] is powered by a high voltage power supply and triggered by signal from timing system. SHBs are derived by independent power supplies with 142.8MHz and 571.2MHz microwave signals. As for buncher and A0 accelerator whose microwave comes from the 1st klystron, any perturbation of power and phase of its exporting microwave also give variation for beam bunching. In other sense, timing stabilization between bunch cells in pre-injector is vital; any variation of them can cause longitudinal jitter. The beam pulse width at gun exit was 1ns FWHM with 1.6ns bottom width. After velocity modulation by SHB1 and SHB2, the beam length was ~900ps and 500ps at the exits without any real acceleration while it was ~60ps and ~10ps at buncher and A0 exit, respectively. Beam energy was about 50MeV at A0 exit[4,6-7] which was measured by an analysis magnet installed at A0 accelerator exit[8]. During bunching process, any variation in longitudinal sequence between pre-injector cells is called longitudinal jitter that may deteriorate beam performance to some extent.

After emission from the gun, electrons were bunched by SHB1, SHB2 and buncher. The current at the exit of the gun and A0 accelerator was measured by two Beam Current Transformers (BCTs); 1st BCT and 2st BCT, which was displayed on oscilloscope at the control room. So we could measure the beam current and its interval between them. Fig. 2 is schematic of the beam current measurement principle.

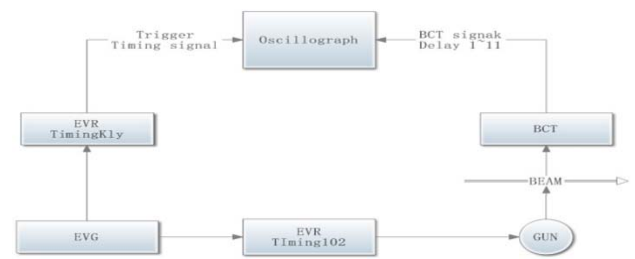

FIG. 2. EVG and EVR measurement scheme

EVG and EVR are event generator and event receiver of timing system, when single beam was bunched and measured as displayed in Fig. 3. The spot on profile monitor produced by analysis magnet was captured as shown in Fig. 4. Then time interval between BCT1& BCT2 was observed and measured in Fig. 5, which was about 21.5ns with tolerance less than 100ps. The calibration coefficient of BCT1 was 10.95, while BCT2 was 9.87. BCT1signal was less than BCT2 signal on oscilloscope, but their actual beam current were about 10A

a)Author to whom correspondence should be addressed.
Electronic mail: muniqbal.chep@pu.edu.pk

and 9A, respectively, therefore, the beam bunch efficiency was about 90%[9].

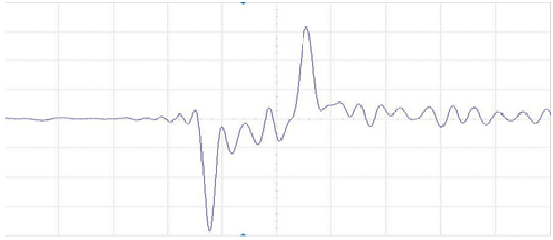
FIG.3. Oscilloscope waveform of BPM measurement

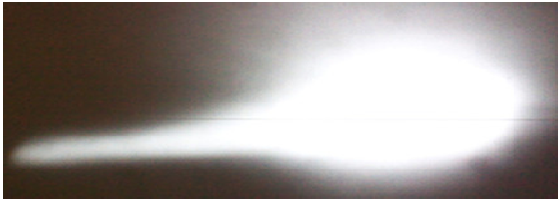
FIG. 4. Beam spot on profile monitor

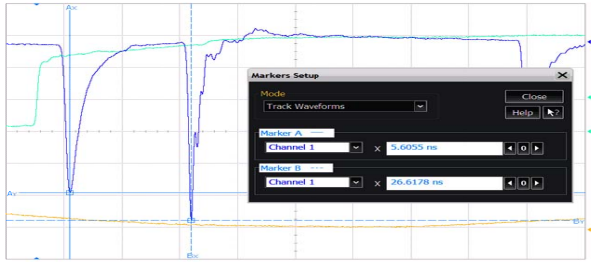
FIG. 5. Oscilloscope waveform of BCT measurement

If there exists instability in pre-injector cells caused due to gun trigger timing jitter[10] or position inconsistency between beam and SHBs or bunch microwave phase, BCT2 signal decreased greatly compared with the normal situation. Fig. 6, is one of jitter situations which describe BCT1& BCT2 unstable. Meanwhile, beam injection rate is decreased or fluctuated.

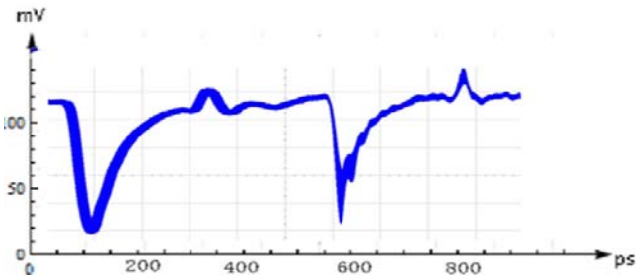
FIG. 6. BCT1 & BCT2 jitters

To study bunch process in pre-injector caused by gun timing jitter in longitudinal direction, simulation was done by PARMELA software at 136.8KeV pulsed beam by adjusting electron trigger timing without changing other parameters. Because the total energy and beam at the Linac exit should be large and stable as possible, the energy spread and emittance is required to be less than $\pm$ 0.5% and 0.1mm.mrad[6], respectively. As, these parameters are strongly dependent on bunching system, so, these parameters are emphasized and analyzed in this work.

In the beginning, gun trigger timing had no longitudinal jitter then, beam parameters at A0 exit were calculated, as shown in Fig 7, which is composed of beam phase spectrum, beam profile, phase spread vs. energy spread and energy spectrum.

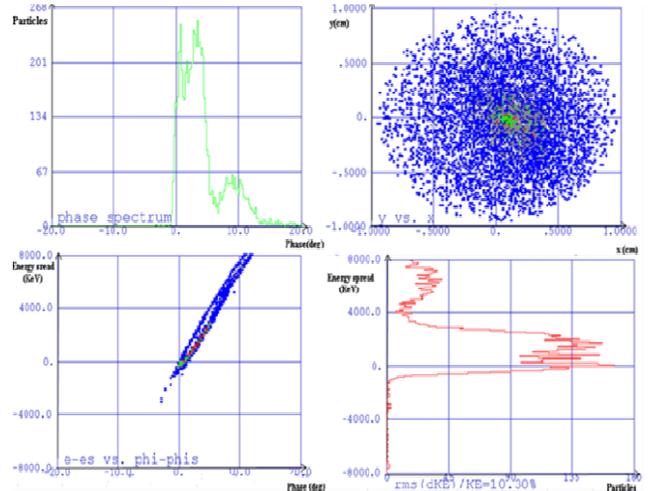
FIG. 7. (Color online). Beam parameters at A0 exit

The values of transverse section, average energy, energy spread, phase of the reference particle and normalized XY emittance were; 1.9cm x 2cm, 47.56MeV 10.30%, 118.15°, & 98.12, 98.76 mm-mrad (3δ), respectively. To compare simulation results, 30ps jitter was considered to calculate beam parameters which are shown in Fig. 8.

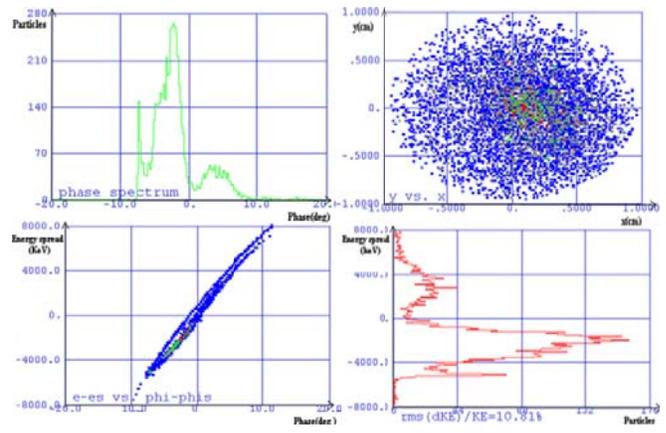
FIG. 8. (Color online). Beam parameters at A0 exit with 30ps longitudinal jitter

The corresponding parameters of transverse section, average energy, energy spread, phase of the reference particle and normalized XY- emittance were; 1.8cm x 2cm, 51.62MeV, 10.81%, 124.39° & 100.70, 21.54 mm.mrad (3δ), respectively. Then, longitudinal jitter of electron gun trigger timing was simulated from 0 ps to 350ps. The beam parameters at A0 exit were calculated and are shown in Figs.9 &10, respectively.

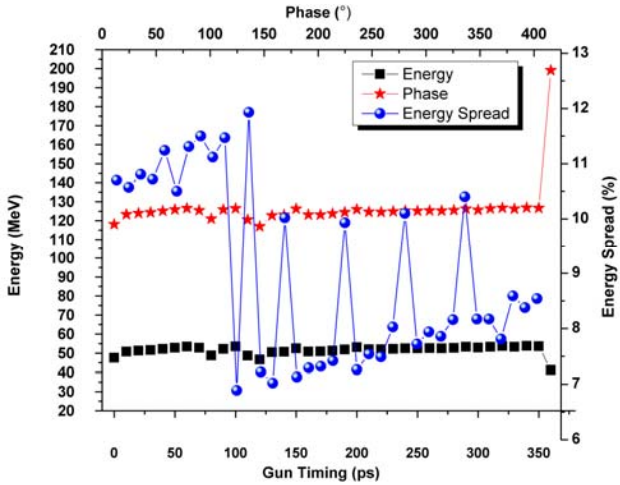

FIG. 9. Beam energy, energy spread and phase at A0

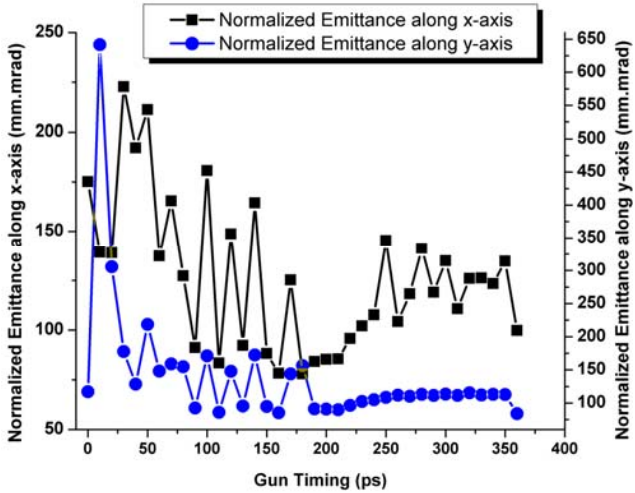

FIG. 10. Beam normalized emittance

Consequently, bunching efficiency from gun to A0 exit on Linac changed greatly by 54.70% (from maximum 83.15% to minimum 28.45%), as shown in Fig 11.

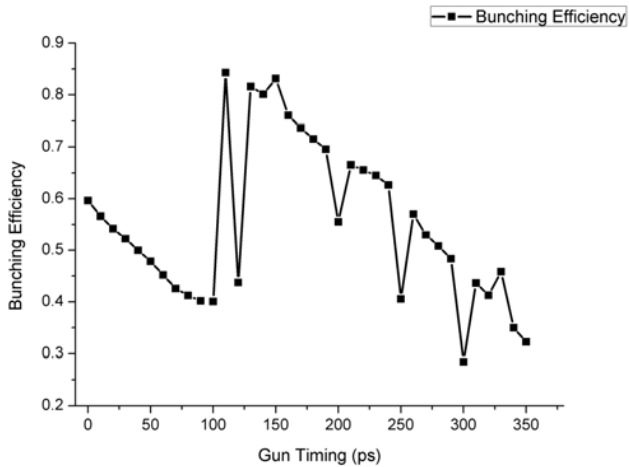

Fig 11: Bunching efficiency

A comparison between maximum and minimum values of beam parameters is given in Table 1.

TABLE 1. Comparison of beam parameters

|  | Beam profile (cm) | Average energy (MeV) | Energy spread (%) | Phase (deg) | Emittance (mm.mrad) | |
|---|---|---|---|---|---|---|
|  |  |  |  |  | X | Y |
| Min | ≤1X1 | 46.60 | 6.89 | 116.99 | 77.91 | 85.16 |
| Max | ≥2X2 | 53.73 | 11.93 | 126.88 | 222.72 | 641.61 |

From simulation results, it is obvious that electron gun trigger timing longitudinal jitter, affects beam performance including the beam size (profile), average energy, energy spread, emittance and phase of reference particle. The simulation results of beam performance are different apparently; the maximum and minimum percentage difference of beam size, average energy, energy spread, phase of reference particle and XY- emittance is; 100%, 13.27%, 42.24%, 7.79% and 65.01%, 86.81%, respectively. Therefore, electron trigger timing is adjusted to optimize electron beam longitudinal position for both good beam performance and storage ring reception for high operation efficiency.

In an injector, it is necessary to control and lessen perturbation of electronic components for good beam performance; the electron gun trigger timing is more possible to be changed due to many invisible factors. In this paper, PARMELA simulations were done by changing gun trigger timing to analyze beam performance which was approximately agreeable to routine operation. It is concluded that any jitter in electron gun will deteriorate beam performance and affect the injection rate. On the other hand, this simulation is helpful for operator to obtain or improve beam performance by optimizing trigger time finely in the bunching process.

[1] L. Mingshan, and M. Iqbal, Rev.Sci. Instrum. **85** 023303 (2014).
[2] P. Shilun, W. Shuhong, High Power Laser & Particle Beams **18** 1349 (2006).
[3] P. Shilun, W. Shuhong, G. Pengda, G. Zheqiao, High Power laser and Particle Beams **16** 795 (2004).
[4] P. Shilun, W. Shuhong, L. Weibin, High Power Lasers and Particle Beams **19** 1537 (2007).
[5] M. Iqbal, A.Wasy, G. Islam and Z. Zusheng, Rev. Sci. Intrum. **85**, 023304 (2014).
[6] P. Shilun, PHD thesis (2006).
[7] C. Yanwei1, W. Shuhong, P. Guoxi, L. Weibin, Y. Qiang, High Energy Physics and Nuclear Physics **30**(6) (2006) 562.
[8] P. Guoxi, Sunyaolin, C. Yunlong,W. Shuhong, Chin.Phys. C **28** (11) (2004) 1214.
[9] W. Shuhong et al., Chin. Physics C **30,** 150 (2006).
[10] T. Asaka, Y. Kawashima, T. Takashima, T. Kobayashi, T. Ohshima, H. Hanaki, Nucl. Instr. and Meth. A 516 (2004) 249.